\documentclass[conference,compsoc]{IEEEtran}
\usepackage{graphicx}
\usepackage{float}
\usepackage{enumerate}
\usepackage{color}
\usepackage{textcomp}
\usepackage{listings, xcolor}

\definecolor{verylightgray}{rgb}{.97,.97,.97}

\lstdefinelanguage{Solidity}{
	keywords=[1]{anonymous, assembly, assert, balance, break, call, callcode, case, catch, class, constant, continue, contract, debugger, default, delegatecall, delete, do, else, event, export, external, false, finally, for, function, gas, if, implements, import, in, indexed, instanceof, interface, internal, is, length, library, log0, log1, log2, log3, log4, memory, modifier, new, payable, pragma, private, protected, public, pure, push, require, return, returns, revert, selfdestruct, send, storage, struct, suicide, super, switch, then, this, throw, transfer, true, try, typeof, using, value, view, while, with, addmod, ecrecover, keccak256, mulmod, ripemd160, sha256, sha3}, % generic keywords including crypto operations
	keywordstyle=[1]\color{blue}\bfseries,
	keywords=[2]{address, bool, byte, bytes, bytes1, bytes2, bytes3, bytes4, bytes5, bytes6, bytes7, bytes8, bytes9, bytes10, bytes11, bytes12, bytes13, bytes14, bytes15, bytes16, bytes17, bytes18, bytes19, bytes20, bytes21, bytes22, bytes23, bytes24, bytes25, bytes26, bytes27, bytes28, bytes29, bytes30, bytes31, bytes32, enum, int, int8, int16, int24, int32, int40, int48, int56, int64, int72, int80, int88, int96, int104, int112, int120, int128, int136, int144, int152, int160, int168, int176, int184, int192, int200, int208, int216, int224, int232, int240, int248, int256, mapping, string, uint, uint8, uint16, uint24, uint32, uint40, uint48, uint56, uint64, uint72, uint80, uint88, uint96, uint104, uint112, uint120, uint128, uint136, uint144, uint152, uint160, uint168, uint176, uint184, uint192, uint200, uint208, uint216, uint224, uint232, uint240, uint248, uint256, var, void, ether, finney, szabo, wei, days, hours, minutes, seconds, weeks, years},	% types; money and time units
	keywordstyle=[2]\color{teal}\bfseries,
	keywords=[3]{block, blockhash, coinbase, difficulty, gaslimit, number, timestamp, msg, data, gas, sender, sig, value, now, tx, gasprice, origin},	% environment variables
	keywordstyle=[3]\color{violet}\bfseries,
	identifierstyle=\color{black},
	sensitive=false,
	comment=[l]{//},
	morecomment=[s]{/*}{*/},
	commentstyle=\color{gray}\ttfamily,
	stringstyle=\color{red}\ttfamily,
	morestring=[b]',
	morestring=[b]"
}

\usepackage[symbol]{footmisc}

\ifCLASSOPTIONcompsoc
  \usepackage[nocompress]{cite}
\else
  \usepackage{cite}
\fi
\begin{document}
\title{Secure Credit Reporting on the Blockchain}

\newcommand{\kw}[1]{{\color{blue}\textbf{\texttt{#1}}}}
\newcommand{\kwp}[1]{{\color{violet}\textbf{\texttt{#1}}}}
\newcommand{\field}[1]{\mbox{\textit{\texttt{#1}}}}

\author{\IEEEauthorblockN{Amir Kafshdar Goharshady}
	\IEEEauthorblockA{IST Austria\\
		Klosterneuburg, Austria\\
		amir.goharshady@ist.ac.at}
	\and
	\IEEEauthorblockN{Ali Behrouz}
	\IEEEauthorblockA{Department of Computer Engineering\\
		Sharif University of Technology\\
		Tehran, Iran\\
		abehrouz@ce.sharif.edu}
	\and
	\IEEEauthorblockN{Krishnendu Chatterjee}
	\IEEEauthorblockA{IST Austria\\
		Klosterneuburg, Austria\\
		krishnendu.chatterjee@ist.ac.at}}

%\author{\IEEEauthorblockN{Ali Behrouz}
%\IEEEauthorblockA{Department of Computer Engineering\\
%Sharif University of Technology\\
%Tehran, Iran\\
%abehrouz@ce.sharif.edu}
%\and
%\IEEEauthorblockN{Krishnendu Chatterjee ~~~~~~~ Amir Kafshdar Goharshady}
%\IEEEauthorblockA{
%	Institute of Science and Technology Austria\\
%	(IST Austria)\\
%Klosterneuburg, Austria\\
%krishnendu.chatterjee@ist.ac.at ~~~~~ amir.goharshady@ist.ac.at}
%}

\maketitle

\begin{abstract}
We present a secure approach for maintaining and reporting credit history records on the Blockchain. Our approach removes third-parties such as credit reporting agencies from the lending process and replaces them with smart contracts. This allows customers to interact directly with the lenders or banks while ensuring the integrity, unmalleability and privacy of their credit data. Most importantly, each customer is given full control over complete or selective disclosure of her credit records, eliminating the risk of privacy violations or data breaches such as the one that happened to Equifax in 2017. Moreover, our approach provides strong guarantees for the lenders as well. A lender can check both correctness and completeness of the credit data disclosed to her. This is the first approach that is able to perform all real-world credit reporting tasks without a central authority or changing the financial mechanisms.
\end{abstract}

\section{Introduction and Preliminaries} \label{sec:intro}
In this section, we first provide a high-level overview of both smart contracts and credit reporting services. Then, we discuss some of the problems that currently exist in real-world credit reporting and argue that these can be mitigated by decentralization and migrating to smart contracts.

\subsubsection*{Blockchain} Blockchain was initially used as a means to achieve global consensus about peer-to-peer cryptocurrency transactions in Bitcoin \cite{nakamoto2008bitcoin}. However, the technology itself is capable of much more than just verifying transactions. Specifically, one can include scripts in transactions, forcing a consensus about the outputs of these scripts. Bitcoin allows simple scripting in a Forth-like loop-free language \cite{bitcoinScript}. A script in a Bitcoin transaction is essentially a program that sets the conditions one must satisfy in order to use the currency units stored in that transaction. For example, a script might ask for a digital signature to gain access to the funds. 

\subsubsection*{Ethereum and Smart Contracts} Ethereum is a cryptocurrency that allows stateful scripts of arbitrary, i.e.~Turing-complete, complexity \cite{wood2014ethereum}. It provides an ecosystem for the development of decentralized applications, called smart contracts, that are executed and verified by the whole Ethereum network. A smart contract can be created by anyone and is stored in a bytecode format on the Blockchain. After its creation, the contract can save data in its own dedicated storage and hold, receive and transfer funds (cryptocurrency units) from/to other people or contracts. It can also interact with other contracts and even create new ones. However, the state and actions of the contract are all controlled by its code and subject to consensus using the Blockchain protocol. After its deployment, one can only interact with a contract by calling its functions which perform actions as programmed by its creator. 
These characteristics, and the inherent lack of a centralized authority in the Blockchain, make smart contracts ideal for implementing a variety of unbreakable financial agreements. For example, a smart contract called BitHalo replaces trusted third-parties and provides escrow services~\cite{bithalo}. We provide another simple example below.

\subsubsection*{Example}
Consider the contract in Figure~\ref{fig:decode}. This contract rewards anyone who can invert a SHA256 hash value. It is written in Solidity which is a widely-used language for programming Ethereum smart contracts and can in turn be compiled to Ethereum bytecode \cite{solidity}.
 The contract creator should provide a value for the parameter \texttt{\_hashed} of the constructor function, which will be stored in the contract. She can also pay some (possibly zero) amount to the contract when creating it. This is signified by the keyword \kw{payable}. After the contract is deployed, anyone can call the function \texttt{claim} and provide an initial value. The contract checks whether this value has the required hash and if so, pays the person who called the function, i.e.~\kwp{msg}.\kwp{sender}, with all the money the contract holds, i.e.~\kw{this}.\kw{balance}. 
 Note that all changes to the state of the contract, along with the messages (function calls) that caused them are stored permanently on the Blockchain and can be read by anyone. Therefore, one can check the contract's balance before attempting to solve the puzzle. Also, after the puzzle is solved, anyone, including the creator of the contract, can read the function call and parameters that led to a solution. This means that while contracts enable us to reach a consensus about the state of a computation, they are not very good at hiding data.

\begin{figure}[H]
\begin{lstlisting}[language=Solidity]
contract HashInvert
{
	bytes32 hashed;
	
	function HashInvert(bytes32 _hashed) payable
	{
		hashed = _hashed;
	}
	
	function claim(bytes32 _initial)
	{
		if(sha256(_initial)==hashed)
			msg.sender.send(this.balance);
	}
}
\end{lstlisting}
\label{fig:decode}
\caption{A Solidity contract that rewards finding a value with a given hash}
\end{figure}

\subsubsection*{Credit Reporting}
A credit report is a document that includes data regarding a person's history of managing credit. This data is collected and maintained by a credit reporting agency and used to assess the creditworthiness of the individual when she applies for new credit. It usually contains the following information \cite{avery2003overview}:
\begin{itemize}
	\item Identifying information, such as the name, address and social security number, of the individual.
	\item Information reported to the credit reporting agency by creditors, such as banks and debt collection agencies, regarding details of current and past loans, leases, credit report requests, utility and medical bills, etc. We refer to each of these as a \emph{credit account}. 
	\item Data collected from public records, such as bankruptcy information.
\end{itemize}

\subsubsection*{Credit Reporting Industry} The companies that compile the credit report are known by various names in different countries. For example, they are called Credit Bureaus in the US and Credit Reference Agencies in the UK. We shall call them CRAs in the rest of this paper. These companies compile credit data and help future lenders decide about extending credit. There are three major CRAs in the US. In 2003, they issued 2 million credit reports each day and each of them was estimated to hold data about roughly 1.5 billion credit accounts belonging to 190 million individuals \cite{avery2003overview}. One of them, Equifax, reported a revenue of nearly \$850 million in the third quarter of 2017 \cite{equifaxrevenue}.

\subsubsection*{Problems with Credit Reporting} It should come as no surprise that the fact that CRAs are collecting and storing vast amounts of sensitive data about hundreds of millions of people is a source of concern for many. To address these concerns, laws and regulations are passed to ensure that the rights of individuals to privacy and fair treatment are not violated. One example is the US Fair Credit Reporting Act (FCRA) of 1970. However, there are still a variety of problems that cannot be addressed by regulation alone. These include:
\begin{itemize}
	\item \emph{Long Update Intervals.} CRAs generally receive information from creditors and other sources once a month and it takes them up to seven days to update the records \cite{avery2003overview}.
	\item \emph{Identification Problems.} The data received by the CRAs does not always include uniquely-identifying information and might be erroneously attributed to the wrong individual \cite{CRaccuracy}. Another related aspect of this problem is identity theft. It was estimated that 12 percent of Americans were victims of identity theft in the 5-year period ending in 2003 \cite{kahn2008credit}. 
	\item \emph{Errors and Inconsistency.} Reports stored by different CRAs can be inconsistent or contradictory \cite{CRaccuracy}. Moreover, it is estimated that as many as a third of all credit reports might contain errors that can lead to denial of access to credit \cite{avery2003overview,golinger1998mistakes}.
	\item \emph{Endemism.} Credit data is usually tied to a single country or jurisdiction. The CRAs cannot access foreign credit information \cite{avery2003overview}. This means that when an individual relocates to a new place, her credit data is effectively erased and her record starts from scratch.
	\item \emph{Data Breaches.} Finally, a major source of discomfort is the possibility of data breaches and unauthorized access to the sensitive credit report information. In a famous catastrophic case in 2017, hackers stole sensitive information about 143 million people in the US from Equifax \cite{ftcwhattodo}.  
\end{itemize}

\subsubsection*{Our Contribution} In this paper, we propose an approach based on smart contracts that can remove the CRAs from the lending process and fixes all the problems mentioned above. In our approach (i)~data updates take only a few seconds, (ii)~identification problems are entirely avoided, (iii)~there is no possibility of inconsistency, (iv)~credit reports can be used globally and (v)~all sensitive information is secured by cryptography. We also give individuals full control over their credit report, allowing them to disclose all or any part of it to others. From the creditors' point-of-view, we guarantee that the report is correct and not editable by its owner and that the creditor can easily check to ascertain that it includes all the data in a requested time-frame.

We now provide a high-level overview of our approach that combines classic constructs in asymmetrical cryptography to achieve secure credit reporting. We first recall the main concepts of encryption, decryption and digital signatures and then proceed with an intuitive description of our method. A more formal treatment is provided in the next sections.  

\subsubsection*{Asymmetrical Cryptography} We assume basic familiarity with asymmetrical and public-key cryptography, as introduced e.g.~in \cite{hoffstein2008introduction}. Formally, we use pairs of keys of the form $(K, k)$ for encryption, decryption and digital signatures. The public key is denoted as $K$ and its corresponding private key as $k$. One can encrypt data using $K$ and then the encrypted data can only be decrypted if one knows $k$. Similarly, one can sign a piece of data using $k$ and this signature is verifiable by anyone who has access to the data and $K$. In particular, a function call in a smart contract always includes the public key $K$ and is signed by the private key $k$. This means that anyone can see the function call data and its caller by reading the Blockchain but no one can make a fake function call on behalf of another person unless they have access to her private key.

\subsubsection*{Underlying Principles of Our Approach} We achieve the guarantees mentioned above by employing a combination of the following techniques:
\begin{enumerate}[(i)]
\item \emph{Identity Management.} We use a decentralized identity management and certification system in which a borrower's identity can be certified by lenders and financial institutions.
 \item \emph{Data Encryption.} We store the credit report data in an encrypted format, using asymmetrical encryption, in a series of smart contracts. The encryption is such that only the owner and creator of a record, or anyone who they authorize by providing the relevant private key, can decrypt it.
 \item \emph{Links Encryption.} We chain the records belonging to each individual in a linked list whose pointers are also encrypted. Hence, not only one cannot read a record without authorization, but it is also impossible to find out to whom a given record belongs or which records belong to a given individual.
 \item \emph{Fraud Prevention.} We use digital signatures and asymmetrical cryptography to avoid fraud. The simplified intuition is that a credit record can be first signed by the creditor and then encrypted using a key pair that is shared with the customer. Then, when another creditor wants to see the record, the customer can decode it and the creditor can check the previous creditor's signature to make sure the customer has not altered the record.
\end{enumerate}

The main novelty of our approach is a combination of these ideas that makes it possible to achieve secure credit reporting on the Blockchain. To the best of our knowledge, this is the first method that can reliably perform all credit reporting tasks without a need for trusted third-parties and centralized authorities, or changing the financial mechanism of credit reporting. 

\subsubsection*{Organization} Our approach consists of three distinct protocols, each realized by a different smart contract. In Section~\ref{protocol:identity}, we present our solution for identity management. Section~\ref{protocol:accounts} is the core of our approach and explains how we handle credit accounts. This is followed by our public records protocol in Section~\ref{protocol:public}. Section~\ref{sec:implementation} provides a short report on a proof-of-concept implementation of our method that is publicly available. We discuss some limitations of our approach in Section~\ref{sec:ext}. Section~\ref{sec:comparison} is a comparison with similar works and finally, Section~\ref{sec:future} concludes the paper with suggestions for future research and development.

\section{Identity Management Protocol} \label{protocol:identity} \label{protocol:name}

One of the main issues in credit reporting, as in many other distributed applications, is identity management. There are two important aspects to this issue: first, one should not be able to masquerade as another person, i.e.~commit identity theft, and second, one should not be able to use more than one identity. Note that in a cryptocurrency setting individuals having multiple identities do not pose a problem, given that this does not entail any benefit. However, in our setting, one person having multiple disjoint credit reports is certainly not acceptable.

A simple solution is to create one or several central authorities that check real-world identities and issue certificates of their validity. This is the solution used, for example, for checking valid HTTPS signatures \cite{durumeric2013analysis}. It is also commonly used for managing the identities of banks, institutions and public authorities. In this paper, we assume that such entities' identities can be verified in this manner.  However, the same approach is not desirable for individual credit customers, given that it puts too much power in the hands of the certificate issuers and they can, at least in theory, bar one from getting access to credit by refusing to issue a certificate.

Our proposal is to let the lenders themselves act as certificate authorities. To be more precise, we allow anyone to issue a certificate verifying the identity of an individual, but we expect the lenders, who are typically banks and financial institutions, to only take into account certificates issued by other banks or institutions that they already trust. Given that the lenders trust data sent by other lenders to the CRAs, which includes identifying information about the owners of credit accounts, it is expected that they agree to accept this same information directly, i.e.~without the CRAs as middlemen, too. While this approach might lead to a situation where a few banks perform most of the certifications, this is not considered to be a problem, since no group of institutions have a monopoly on certification and every lender who is willing to extend credit to an individual can also certify her identity.

\subsubsection*{Data Fields of the Identity Management Contract} We now describe our identity management protocol more formally. Our approach is realized by a single instance of a smart contract that keeps track of every individual by storing the following data:
\begin{itemize}
	\item The \emph{public key} used by the individual.
	\item \emph{Fingerprint.} A unique identifier that can be used in real world to check the individual's identity. This can be biometric data or any other data that is unique to the individual. Our approach is not dependent on the exact standard that is used for creating fingerprints, but they should be standardized. If this data is sensitive, one can store a hashed version of it. For example, we can use a hashed version of the individual's country of nationality, appended with her national identification number (or social security number in case of the US).
	\item \emph{Two Pointers.} A pointer to the first public record of the individual and another one to her first credit account. These will be discussed in more detail in the next sections.
	\item \emph{Certificates.} A list of public keys of individuals or institutions who have verified this identity in the real world. 
\end{itemize}

\begin{figure}[H]
	\resizebox{\linewidth}{!}
	{
		\includegraphics{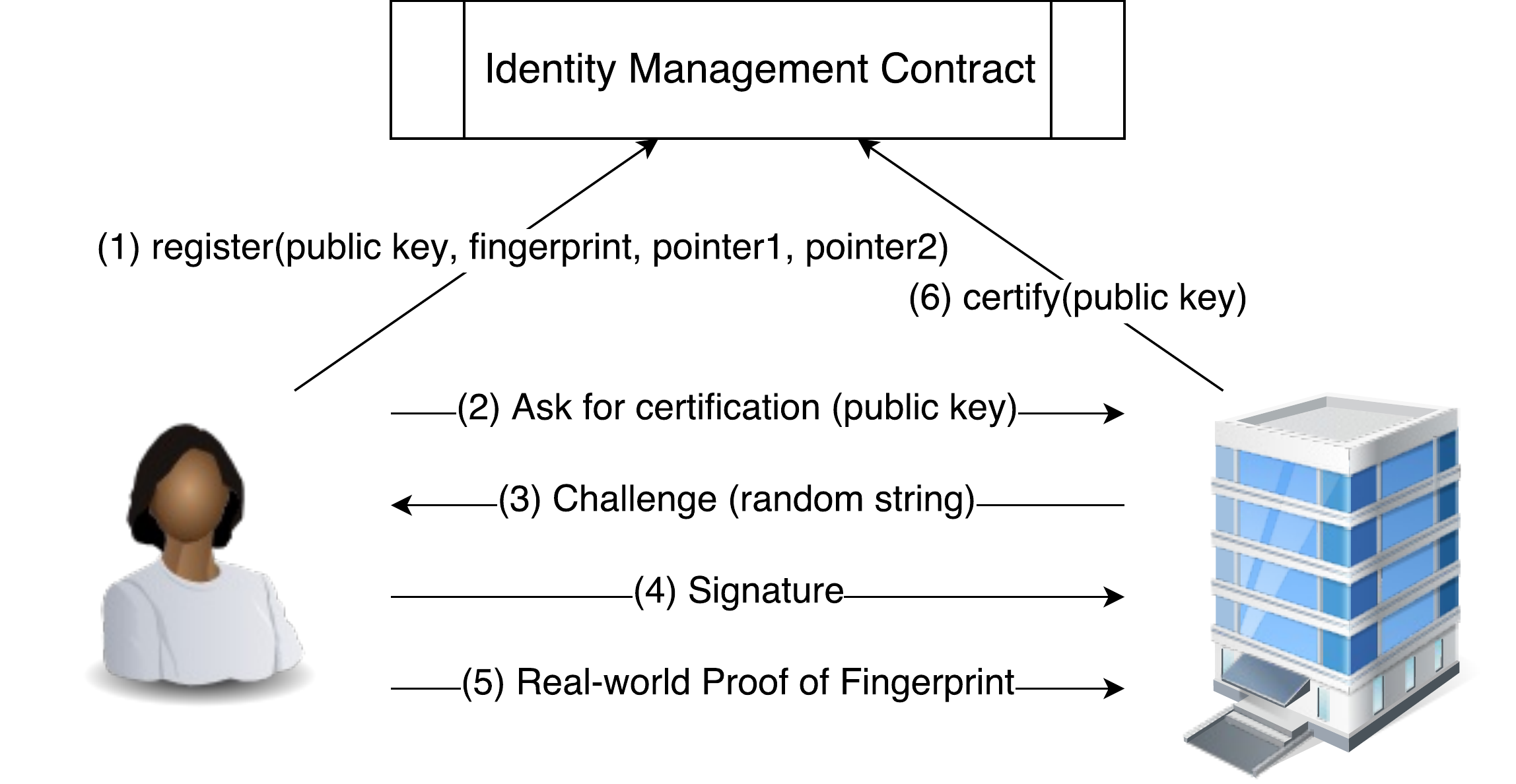}
	}
	\caption{Interactions between an individual, an institution and the identity management contract. Numbers denote the order in which the actions are taken.}
		\label{fig:id}
\end{figure}

\subsubsection*{Functions of the Identity Management Contract}
We now describe how our identity management smart contract works. This is summarized in Figure~\ref{fig:id}. Anyone can register in this contract by calling the \texttt{register} function and providing her own desired public key and (possibly fake) fingerprint. The contract even allows several public keys to be registered as corresponding to the same fingerprint. After a public key and its corresponding fingerprint are added to the contract, anyone can call the function \texttt{certify} and announce that they have checked an identity in the real world and would like to certify it. In this case, the caller's public key is added to the list of certificates. There is also a \texttt{decertify} function that can be used to revoke the certification.

\subsubsection*{Safety against Sybil Attacks}Effectively, one can create as many fake identities and certify them with as many self-created keys as she wishes. However, the lenders would only consider certificates from other trusted lenders or institutions. Such an institution would (i)~ask the individual to sign a random piece of data using the private key corresponding to the desired public key to ensure that she has access to it, (ii)~require real-world verification of the fingerprint, and (iii)~require that no other public key is already certified as corresponding to the same fingerprint by another trusted institution. Only when all of these conditions are met would the institution certify the identity. 

\subsubsection*{Legal Guarantees} Note that the institutions, such as banks, have publicly announced public keys and will be subject to legal action, under FCRA or similar regulation in other countries, should they provide false certifications or decertifications. The process is also uniquely transparent, given that all changes to the contract are permanently recorded in the Blockchain. An individual can ask each lender she deals with to certify her identity so that the respective credit account is also trusted by future lenders.

\subsubsection*{Privacy} Our protocol preserves user privacy. The fingerprint is associated with a public key that does not appear in the credit accounts, ensuring that even having access to a person's fingerprint cannot be used to extract information about their non-public credit records. In the next sections, we will show that an attacker with access to the Blockchain cannot read data about the records, such as account details, and is even unable to infer the owner of a given record.

\section{Credit Accounts Protocol} \label{protocol:accounts}

We now turn to the core of our approach, which is a protocol for storing credit accounts' data. We introduce a smart contract for modeling credit accounts. Each account is realized by one instance of this contract. This is in contrast to Section~\ref{protocol:identity} where all identities were stored in a single instance of the identity management contract.

As mentioned earlier, we rely on asymmetric (public-key) cryptography. To achieve the desired level of security, we will introduce several new keys in this section. Therefore, to avoid confusion, we use the term ``true identity'' to refer to the key pair which is publicly known to belong to an institution. Similarly, an individual's true identity is the key pair with which she registers in the identity management protocol and for which she obtains certificates. Also, we use upper-case $K$ for public keys and lower-case $k$ for private keys.

We store a singly linked list of each individual's credit accounts, with each account providing a pointer to the next. Note that in Ethereum each deployed instance of a smart contract is uniquely addressable and therefore these pointers are well-defined. The identity management contract provides a pointer to the first credit account. Moreover, these pointers are encrypted, as explained below, and hence they can only be traversed if the individual owner allows it.

\begin{figure}[H]
	\resizebox{\linewidth}{!}
	{
			\includegraphics{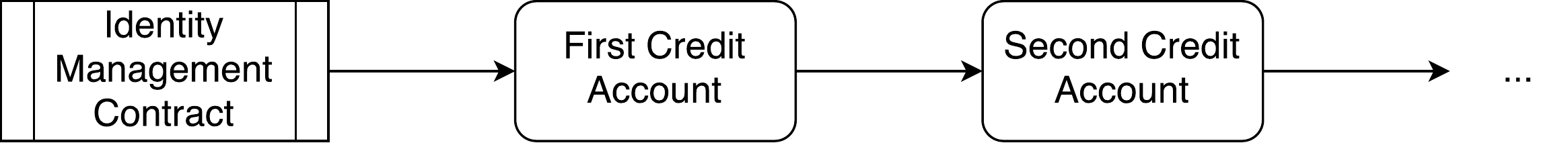}
	}
\caption{Each credit account is stored in its own instance of the credit account contract. The arrows denote encrypted pointers.}
\end{figure} 

We now proceed to define the data stored in a credit account contract and the process for its creation, management and use as part of a credit report.

\subsubsection*{Key Generation}
\begin{figure}[H]
	\begin{center}
	\resizebox{!}{2.9cm}
	{
		\includegraphics{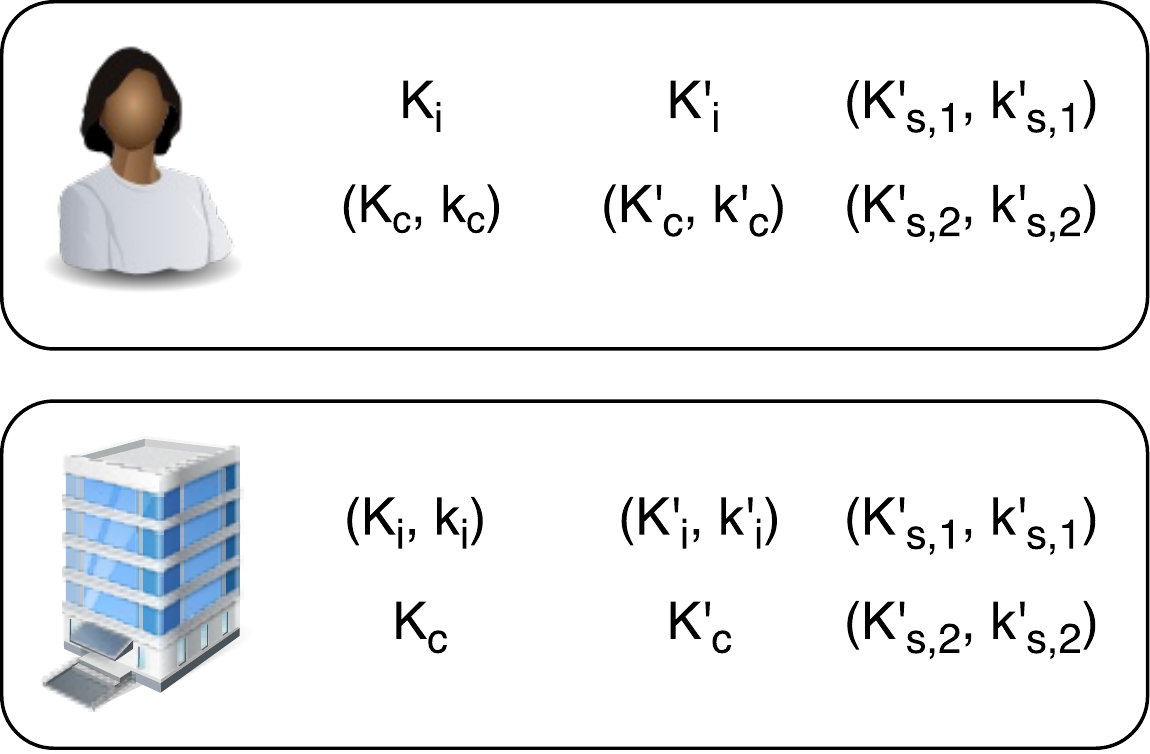}
	}
	\end{center}
\caption{Key distribution prior to deployment of a Credit Account Contract}
\label{fig:key-distro}
\end{figure}

Let the institution's true identity be $(K_{i}, k_{i})$ and the customer's true identity $(K_c, k_c)$. When, after verifying a customer's identity and credit record, an institution agrees to extend credit to a customer, they ask her to create a new key pair $(K'_c, k'_c)$, called customer's account-specific keys. The institution in turn creates its own account-specific keys $(K'_i, k'_i)$. Then, each side provides the other side with their account-specific public key. Finally, they create and fully exchange two other pairs of keys $(K'_{s,1}, k'_{s,1}), (K'_{s,2}, k'_{s,2})$, which we call account-specific shared keys. Hence, the keys are distributed as in Figure~\ref{fig:key-distro}.

\subsubsection*{Contract Creation} 
\begin{figure}[H]
	\begin{center}
		\resizebox{\linewidth}{!}
		{
			\includegraphics{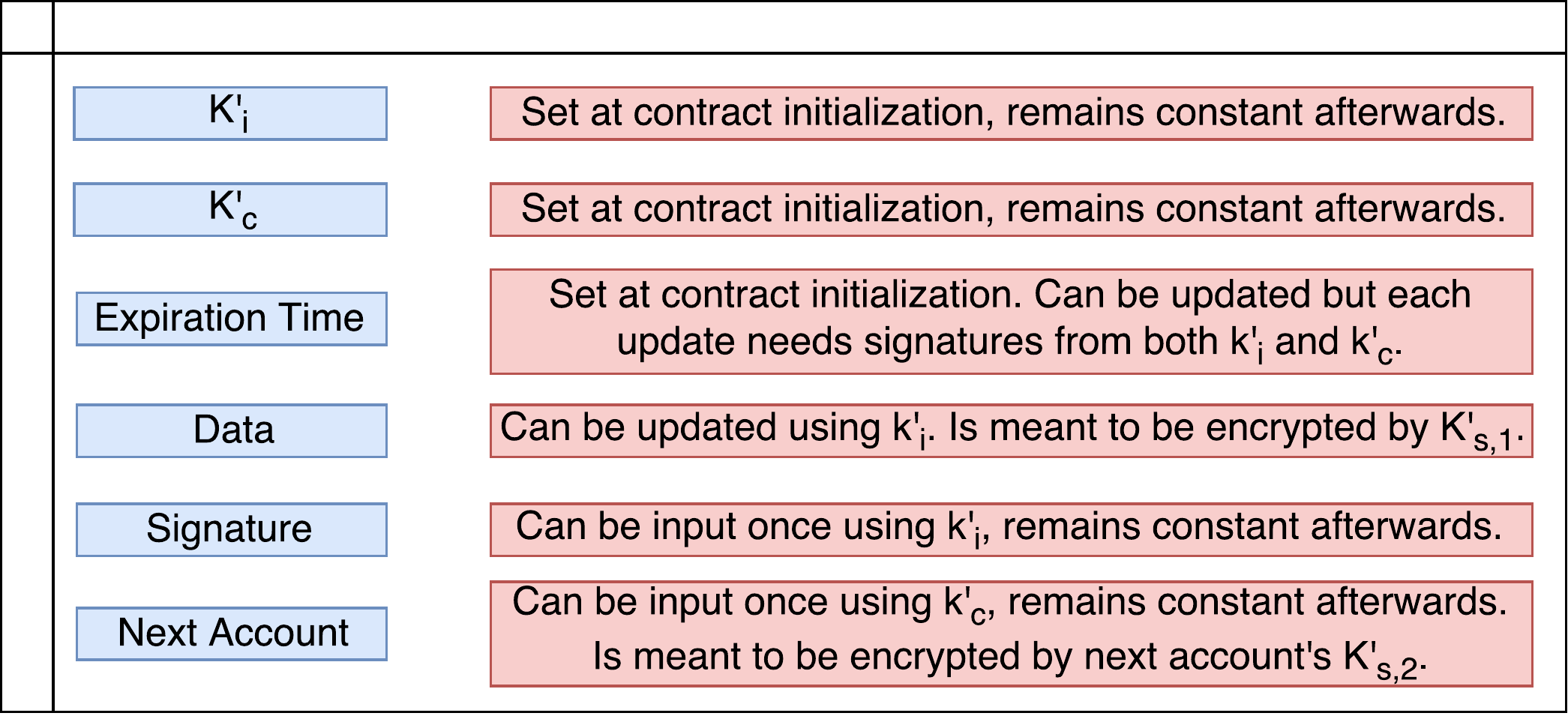}
		}
	\end{center}
	\caption{Data fields and constraints in a Credit Account Contract}
	\label{fig:ct}
\end{figure}

At this point, the institution creates a new instance of the credit account contract and publishes it on the Blockchain. Figure~\ref{fig:ct} shows the data stored in this contract and the conditions enforced by the contract for changing this data. The contract stores public keys of the customer and the institution, i.e.~$K'_c$ and $K'_i$. These are set at the beginning and are not changeable afterwards. Note that the contract does not store true identities, but uses contract-specific public keys instead. All function calls are also performed using contract-specific keys. The reason behind this is that anyone has access to the data stored on the Blockchain and one must not be able to read the true identities using publicly available data. The contract also has an expiration time which can be changed only if both parties agree on the new value. 

\subsubsection*{Commitment} After the contract is deployed, both parties must commit to it by verifiably connecting it to their true identity. The institution does this by signing the contract address, $K_i$ and $K_c$ using its true identity and adding the signature to the contract. This signature cannot be changed after it is added. At this point, the customer can check the signature. If the check passes, she adds the contract to her record by letting her last account's \field{Next Account} field point to this contract by storing its address encrypted using $K'_{s, 2}$. Note that the \field{Next Account} field can be changed only once and hence the contract cannot be removed from the customer's report when added. The institution can now check that the contract is added to the linked list of customer's report using $k'_{s, 2}$.

\subsubsection*{Credit Report Data} Finally, the institution can change the contents of the field \field{Data} as long as the expiration time has not passed. It can store all the relevant data about this account that should appear in a credit report. This data is always encrypted using $K'_{s, 1}$ and is hence accessible to both the institution and the customer, who know $k'_{s, 1}$, but not to anyone else. 

Note that we are assuming this data fits in a single transaction of the underlying Blockchain and can hence be changed by a single function call. This is because, to the best of our knowledge, most credit account reports contain only a few lines of data. However, this assumption does not affect the generality of our approach. If the data happens to be too big, one can store it in an external service, such as IPFS, which is a peer-to-peer network for file storage and transfer that supports immutable version control using a structure very similar to the Blockchain \cite{ipfs}. Then one can fill the \field{Data} field with an address/identifier of the original data in IPFS.

\subsubsection*{Reading a Credit Report} When another institution wants to read customer data, it would need the values of $k'_{s, 2}$ for each of the contracts to be able to decrypt the links and traverse the linked list.\footnote[1]{Alternatively, the customer can provide the decrypted contents of the \field{Next Account} fields and the public keys $K'_{s,2}$. The institution can then verify the correctness of this data by encrypting them using these keys and checking that they lead to the same encrypted values that are saved in the contracts.} These can only be provided by the customer. Hence, one cannot find out which accounts belong to an individual, unless that individual allows access. When access is granted, the institution can easily find out when it reaches the end of a report given that the \field{Next Account} field is only empty at the end of the linked list. The institution can also see the beginning time of a contract by looking up the number of the Blockchain block where the contract was first created. Expiration times of the credit accounts are publicly visible on the Blockchain, but not their data. Should the customer decide to allow the institution to read a contract, she can provide them with the contract-specific $k'_{s, 1}$ to access the \field{Data} field\footnote[7]{As in the previous case, an alternative is to provide the decrypted contents of \field{Data} and the public key $K'_{s,1}$.} and with the lender's true identity, $K_i$, to verify the signature.

Note that an individual can add as many credit accounts as she wishes to her linked list, acting as both the institution and the customer. This can be used to initialize the linked list by an account when creating an identity, and also to resist any attempt by an institution to find out the true number of accounts belonging to an individual.

\section{Public Records Protocol} \label{protocol:public}

Our protocol for storing public records is similar to the one we described for credit accounts. However, in this case the protocol becomes much simpler, because unlike credit accounts, public records can be made without the consent of their individual owners. For example, a court does not need permission from an individual to add a declaration of her bankruptcy to her credit report.

Similar to the previous section, we store public records in a singly linked list. Each record is an instance of the public record contract. As previously mentioned in Section~\ref{protocol:name}, there is a pointer from an individual's identity to her first public record, which can be created by herself.

Unlike credit accounts, the pointers used to connect public records are not encrypted. This allows anyone to follow the list of public records corresponding to an identity. Moreover, anyone can add a new public record to the end of any of these lists. This is not problematic, given that lenders will only take the records issued by real public institutions into account. Simply, each record is either added using an unknown identity, in which case it can be considered as spam and ignored\footnote{Producing spam is not free given that one has to pay for its gas fees. This is the native Ethereum solution to combat spam and it naturally extends to our contracts. On the other hand, when reading the records, one can differentiate spam entries pretty fast, by simply checking the identity of their signatures. Note that reading the blockchain is free but writing to it is not.}, or by an official identity, in which case it is either correct or can be corrected by the same authority. Again, note that all changes to the contracts are permanently saved on the Blockchain and that official authorities are bound by legal responsibilities and cannot simply issue false records.

\begin{figure}[H]
	\begin{center}
		\resizebox{\linewidth}{!}
		{
			\includegraphics{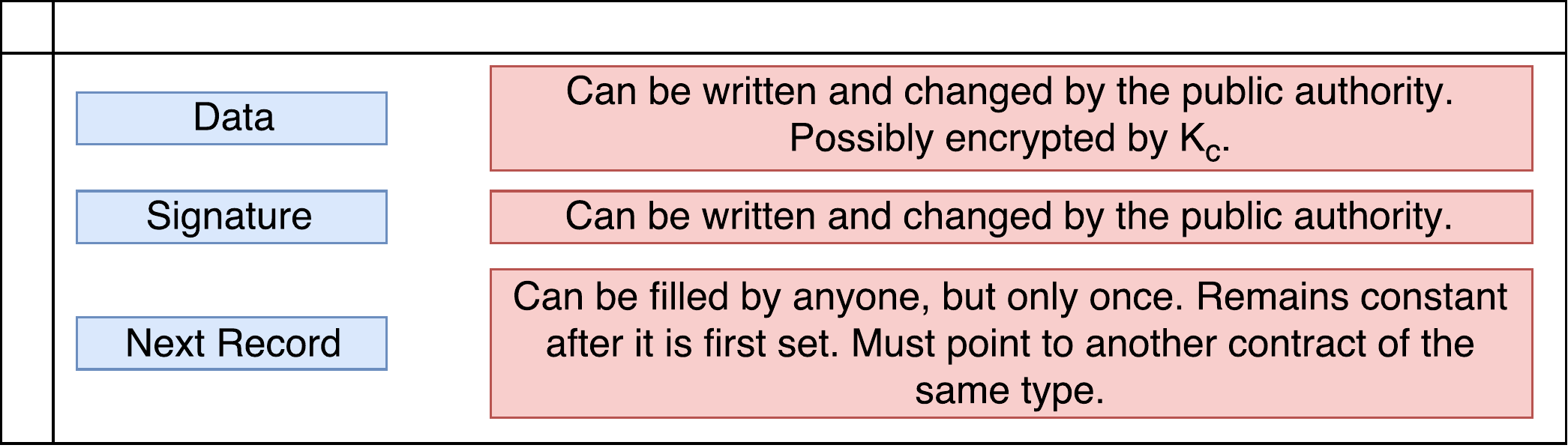}
		}
	\end{center}
	\caption{Data fields and constraints in a Public Record Contract}
	\label{fig:pr}
\end{figure}

We now define the structure of our public record contract more formally. Figure~\ref{fig:pr} shows the data fields in a public record contract together with the constraints enforced by the contract. 

\subsubsection*{Contract Creation}
The public authority creates an instance of this contract and publishes it on the Blockchain. The authority has access to the individual's fingerprint and can hence add the record to linked lists corresponding to all identities that have that fingerprint. To do so, the authority follows the \field{Next Record} pointers until it reaches the end of the linked list, and then sets the final \field{Next Record} to point to the new instance of the contract. Note that anyone can set the value for \field{Next Record}, the only limitations are that (i)~it can be filled only once and (ii)~it must keep the linked list valid and extensible. We refer to the latter condition as ``validity''.

\subsubsection*{Credit Report Data}
The other two data fields in this contract, \field{Data} and \field{Signature}, are under complete control of its issuer. \field{Data} is meant to contain any relevant information that should be considered part of the credit report. The authority can decide whether to fill this data without encryption, hence allowing public access to it, or encrypt it using $K_c$, so that it is only accessible by the individual owner herself. In the latter case, the authority signs the original unencrypted \field{Data} and stores this signature in the contract. This ensures that the individual owner can both read and prove what is saved in \field{Data} and is the only person, other than the public authority, who can perform these actions. 

\subsubsection*{Reading a Credit Report}When an institution decides to read the public records of an individual, it simply follows the linked list, ignoring any entries created by unknown identities. In case it faces an encrypted entry by a trusted public authority, it asks the individual owner to decrypt the \field{Data} field and provide the decrypted text. It then checks the signature to make sure that the text was not changed by the owner. 

\subsubsection*{Importance of Validity} Except for the validity condition, all other aspects of our public records protocol are simplifications of those used for credit accounts. When dealing with credit accounts, the pointers used for our linked list were filled by the individual who owned them and there was no fear that she might intend to destruct the whole linked list. Also, the signatures provided by the institutions guaranteed that one cannot add another person's record to her linked list without getting caught. However, in the case of public records, anyone can add a new element to the linked list and fill the \field{Next Record} fields. These fields remain immutable after they are first filled. So, a natural attack would be to fill them with invalid pointers, i.e.~pointers that do not hold the address of a valid contract of the same type. This will make it impossible for others to keep adding records. Another malicious behavior is adding the same instance of a record to the linked lists belonging to two different individuals. This will effectively merge the two lists. It is therefore of utmost importance to keep the linked lists valid.

\subsubsection*{Enforcing Validity} To avoid the attacks described above, we do not allow the individuals to create instances of our public record contract directly. Instead, we develop a so-called ``factory'' contract that can be called by anyone to create valid instances of the public record contract. The factory contract also keeps track of the addresses of all valid public record contracts instantiated using it and whether they have been added to a linked list. On the other hand, each such instantiated contract includes an immutable pointer to the parent factory contract. When a new contract is being added to the linked list, it is first checked against the factory contract to ensure it respects validity. 

\subsubsection*{Checks for Adding a New Entry} Formally, when an individual attempts to set a value for the \field{Next Record} pointer, the public record contract performs the following actions:
\begin{enumerate}
	\item It first checks the value with its own parent factory contract. If the provided value is not a valid address or if it does not point to a contract created by the same factory, the operation is rejected.
	\item It checks that the person (public key) trying to add the new record is the same person who created this new record. One cannot add records authored by others to the linked list. 
	\item It queries the parent factory contract to make sure this is the first time the given record is being added to a linked list. The parent factory contract remembers this query and will answer negatively to any following queries about the same record.
	\item It checks that the new record has an empty \field{Next Record} field. This is equivalent to checking that only a single record is being added.
\end{enumerate}
These checks ensure that the linked lists remain valid and accessible to everyone for adding new entries.

\noindent\textbf{Deanonymization.} The fact that public records are not encrypted means that they can be used to deanonymize users. For example, public records of bankruptcy often include names of individuals and their national identity numbers, which might be the same as fingerprints. However, the only additional  data that can be inferred by such deanonymization is the individual's public key $K_c$. As mentioned before, this key is not saved in any of the credit account contracts and cannot be used to infer any non-public information about the individual. Note that the public records themselves are, and should be, accessible to everyone.   

\section{Implementation} \label{sec:implementation}

We have implemented our approach in Solidity to demonstrate the feasibility of the ideas and structures that we suggest. A proof-of-concept implementation, together with instructions for its deployment and testing, is available at \texttt{pub.ist.ac.at/\texttildelow akafshda/credit-reporting}.

Our implementation is entirely loop-free and all of its function calls terminate after executing a small (constantly-bounded) number of instructions. Hence, our gas cost, i.e.~the cost one must pay for execution of commands in Ethereum smart contracts~\cite{wood2014ethereum}, is very little.

\section{Limitations} \label{sec:ext}

In this section we discuss some of the limitations of our approach and ideas for addressing them.

\noindent\textbf{Inherited Limitations.} The goal of our approach is to remove the CRAs from the credit reporting process, allowing the same financial mechanisms that are currently established to run without relying on a middleman or trusted third party. This means that our approach essentially inherits any limitation of the traditional centralized credit reporting that is not due to the CRAs. In particular, if an individual has two (or more) provable identities in the real world, e.g.~two distinct names and national identity numbers, then she can sign up in our identity management contract twice and obtain certificates for both identities. Note that this attack is not dependent on the existence or lack of CRAs and is also possible under the current credit reporting systems that have CRAs. Moreover, migrating to the blockchain cannot solve this problem given that the smart contracts can only access the data saved on the blockchain and have no way of realizing that the same person has fake or multiple identities in the real world.

\noindent\textbf{Cryptographic Primitives.} The security of our approach is dependent on the security of the cryptographic primitives that are used. Users must keep in mind that any data that is saved on the blockchain is permanent and cannot be deleted. In several protocols explained above, data encryption is used in order to hide the data and restrict public access to it. If/when the underlying cryptographic primitives are broken, this data can be recovered. Therefore, it is advisable to refrain from saving the actual credit data in smart contracts or IPFS, but instead rely on saving its hash. This way the data would be provable, but cannot be obtained even if the cipher breaks in the future. The downside to this method is that the individual has to keep safe copies of all the actual credit report data and can only use our approach for proving its correctness. 

\noindent\textbf{Legal Problems.} Our approach does not intend to address legal aspects of credit reporting. We provide a solution that works under minimal legal assumptions, i.e.~prohibition of fraud. Wrong information provided by an institution can be traced back to their originator who in turn has the ability to fix them. Problems arising due to inconsistencies in laws and regulations, especially in a multi-jurisdiction environment, are beyond the scope of this paper.

\section{Comparison with Related Works} \label{sec:comparison}

In recent years, a variety of financial agreements and processes have been implemented using smart contracts. The first such contract was BitHalo \cite{bithalo}. It replaced middlemen in an escrow, and allowed distrusting parties to buy and sell goods over the internet with security and peace of mind. Unfortunately, it was commonly used in darknet markets such as the Silk Road \cite{silkroad}. Another notable example is the concept of decentralized autonomous organizations \cite{vigna2016age}. These are organizations that are entirely governed by rules written as smart contracts.

After the Equifax breach \cite{ftcwhattodo}, which led to a leak of sensitive data belonging to more than 140 million people, several authors suggested that credit reporting can potentially benefit from decentralization and Blockchain techniques \cite{floyd,huffpost}. However, no concrete approach was introduced to achieve this goal. We filled this gap in this paper by introducing a simple smart-contract-based approach for credit reporting.

At the same time that we were developing our approach, a startup, called Bloom, was created to perform credit scoring on the Blockchain \cite{bloom}. The full details of their protocol is not published and their code is under development. We are not aware of the exact extent of similarity between our approaches. However, based on the Bloom whitepaper \cite{bloom}, there seems to be several fundamental differences.

Our approach provides the exact same financial mechanisms as real-world credit reporting and our goal is to simply remove the CRAs from the process and migrate to the Blockchain, while keeping everything else intact. In contrast, Bloom modifies the financial principles of credit scoring with the goal of making credit accessible to a wider population. It defines its own credit score and argues for its adoption. This score depends not only on the credit history, but also on heuristics such as the graph of acquaintances of a borrower and whether they are willing to vouch for her creditworthiness.

Another main difference is the role of laws and regulations. We assume that all institutions are bound by regulations such as the FCRA and hence the fact that all their actions are provably recorded on the Blockchain is a guarantee that they will not provide false data, and even if they do, they will be subject to legal action and the data can be corrected. In contrast, in line with its goal of making credit more accessible, Bloom opts for a method whose goal is to allow even anonymous lenders and borrowers to take part. This is considerably different from the current status of credit reports where the banks and financial institutions only take reports from other comparable institutions into account.

Finally, Bloom is susceptible to Sybil attacks, where an attacker fakes many identities and keeps giving loans to herself, therefore increasing her creditworthiness. According to its current whitepaper, the solution for avoiding this attack is having several ``trusted participants'' who ``will be manually vetted by the Bloom team''. This gives undue advantage to the Bloom team and is effectively equivalent to having Bloom as a third-party instead of the CRAs. In contrast, a Sybil attack does not entail any benefit in our approach.

\section{Conclusion} \label{sec:future}
In this paper, we presented the first solution and a basic prototype for performing secure credit reporting with no third-parties. In Section~\ref{sec:intro} we identified five problems with current systems of credit reporting that can be avoided by migrating to the blockchain. We review how our approach solves these problems:
\begin{itemize}
	\item \emph{Long Update Intervals.} Each update to the credit data is done via a single function call in one of the smart contracts. Hence, it takes a few seconds to be added to the Ethereum blockchain, and after a few minutes one can be sure that it will not be reverted.
	\item \emph{Identification Problems.} The certification procedure presented in Section~\ref{protocol:identity} ensures that only valid real-world identities will be trusted by the institutions and that each real-world identity can only be represented by a single public key $K_c$.
	\item \emph{Errors and Inconsistency.} Inconsistency can only be caused by forks in the blockchain and disappears as soon as the fork is resolved. In our contracts, erroneous data added by an institution can always be fixed by the same institution\footnote{Note that while the contents of the blockchain are immutable, the values of contract variables are not. The blockchain saves the sequence of changes to these values. Hence, once an error is fixed, its history remains in the blockchain.}. Moreover, the source of such data can be provably ascertained. Hence, the institutions are legally bound to fix it.
	\item \emph{Endemism.} Using the Ethereum blockchain, the contracts and their data can be used in the same manner all over the world.
	\item \emph{Data Breaches.} There is no central authority controlling all the credit report data. Moreover, each credit account is secured by its own dedicated keys. Hence, a large-scale breach is impossible unless the underlying cryptographic ciphers break.
	 
\end{itemize}

There are several directions for future research and development. A first step is creating a more user-friendly interface, especially one that abstracts away the underlying cryptography. Another interesting problem is to run real-world large scale experiments to see if creditors and borrowers feel comfortable with this new approach to credit reporting. On the theoretical side, an interesting problem would be to incorporate multi-party computations in a way that a creditor does not need to read the credit report of an individual directly, but can instead rely on a process that, using data provided by the individual and the creditor and a secure connection to the Blockchain, decides whether the individual satisfies specific credit requirements set by the creditor and if so, produces an unforgeable certificate of her creditworthiness. 

\section*{Acknowledgments}
We are thankful to the reviewers for raising points that significantly improved this article. The research was partially supported by Vienna Science and
Technology Fund (WWTF) Project ICT15-003, Austrian Science
Fund (FWF) NFN Grant No S11407-N23 (RiSE/SHiNE) and ERC
Starting grant (279307: Graph Games). The first author is supported by an IBM PhD Fellowship.
\newpage

\bibliographystyle{IEEEtran}
\bibliography{refs}

% that's all folks
\end{document}